\UseRawInputEncoding
\documentclass[12pt]{article}
\usepackage{graphicx}

\begin{document}

\begin{center}
\Large {\bf  Optical equations for null strings}
\end{center}

\bigskip
\bigskip

\begin{center}
D.V. Fursaev
\end{center}

\bigskip
\bigskip

\begin{center}
{\it Dubna State University \\
     Universitetskaya str. 19\\
     141 980, Dubna, Moscow Region, Russia\\

  and\\

  the Bogoliubov Laboratory of Theoretical Physics\\
  Joint Institute for Nuclear Research\\
  Dubna, Russia\\}
 \medskip
\end{center}

\bigskip
\bigskip

\begin{abstract}
An optical equation for null strings is derived. The equation is
similar to Sachs' optical equations for null geodesic congruences. The string optical equation is given in terms of a single
complex scalar function $Z$, which is a combination of spin coefficients at the string trajectory. Real and imaginary parts of $Z$
determine expansion and rotation of strings.  Trajectories of strings can be
represented by diagrams in a complex $Z$-plane.  Such diagrams allow one to draw some universal features of null strings in different backgrounds.
For example, in asymptotically flat space-times $Z$ vanishes as future null infinity is approached, that is, gradually shapes
of strings are ``freezing out''.  
Outgoing gravitational radiation and flows of matter
leave ripples on the strings.  These effects are encoded in subleading terms of $Z$. String diagrams are demonstrated
for rotating and expanding strings in a flat space-time and in cosmological models.
\end{abstract}

\newpage

\section{Introduction}\label{intr}

One-dimensional objects, strings, which move with the speed of light have been studied since 1970's. 
Points of null strings move  along  trajectories of light rays, orthogonally to strings. Null strings were introduced by Schild \cite{Schild:1976vq}
as microscopic objects in the theory of strong interactions.  Later on such strings attracted much attention
as a tensionless limit of the string theory,
since they may capture different features of fundamental strings at Planckian energies \cite{GM1}, \cite{GM2}.  Modern scenario
how fundamental tensionless
strings may emerge in a quantum gravity theory have been discussed in \cite{Lee:2019wij},\cite{Xu:2020nlh}.
 
If fundamental tensionless strings were produced in the early universe they might be stretched to cosmological scales and became
cosmic strings.
At the present moment, such mechanisms are known for fundamental strings with a finite tension, tensile strings \cite{Sarangi:2002yt}, \cite{Copeland:2003bj}.  

Cosmic tensile strings  \cite{Kibble:1976sj}, \cite{Vilenkin:2000jqa}  have a finite rest mass per unit length.  Tensionless strings have zero
rest mass but a finite energy per unit length. So one can also call  these two types of cosmic strings, respectively, massive and massless strings
\cite{Fursaev:2017aap},\cite{Fursaev:2018spa}.
These names
are more appropriate in studying physical effects caused by the gravitational field of the stings on the surrounding matter.

Massless cosmic strings in a flat space-time can be obtained from 
massive cosmic strings as a limiting case,  when the velocity of the string reaches the speed of light, mass tends to zero, while energy  remains finite \cite{Barrabes:2002hn}. As a result of this limit, a holonomy along a closed contour around a massive string 
is transformed into a non-trivial holonomy  \cite{vandeMeent:2012gb} which belongs to the parabolic subgroup
associated to null rotations. Effects caused by massless strings look as mutual transformations
of trajectories of massive bodies or light rays, when the string moves in between two trajectories.

A method how to describe physical effects around massless cosmic strings 
has been developed in \cite{Fursaev:2017aap}, \cite{Fursaev:2018spa}  for strings in flat and de Sitter space-times.
Here backreaction effects can be described analytically due to maximal isometries. For some extension of these results see  
\cite{Fursaev:2020oom}. 

Massless cosmic strings generate perturbations of the velocities of bodies resulting  in overdensities of matter. 
The strings also shift energies of photons, and may yield additional 
anisotropy of cosmic microwave background. These effects are direct
analogs  of, respectively,  wake effects \cite{Brandenberger:2013tr} and the Kaiser-Stebbins effect \cite{Stebbins:1987va}, \cite{Sazhina:2008xs} 
known for tensile cosmic strings.  Thus, if cosmic strings which move with velocity of light do exist in the Nature their physical effects 
can be discovered in future cosmological and astrophysical observations.

In the rest of this paper we consider general properties of trajectories of null strings introduced in \cite{Schild:1976vq}. 
We use name ``null strings'' instead of ``massless strings'' to keep connection to earlier publications.  As we will see, the massless strings
in space-times with parabolic isometries studied in  \cite{Fursaev:2017aap}, \cite{Fursaev:2018spa}, \cite{Fursaev:2020oom}
are a subclass of the null strings.

Solutions to equations of motion for null strings in various gravitational backgrounds have been 
presented in many publications, see e.g. \cite{Kar:1995br} - \cite{Dabrowski:2002iw}. The solutions 
are coordinate and parametrization dependent.  To extract a useful physical information one needs
some invariant characteristics of the string trajectories (world-sheets).

The aim of the present work is to identify such characteristics  and establish universal 
features of string trajectories. We treat them  as one-parameter null geodesic congruences and derive equations 
analogous to Sachs' optical equations. The string optical equations are given in terms of a single
complex function $Z$ which determines expansion and rotation of the string.  

The paper is organized as follows. We start in Sec. \ref{def} with description of trajectories of string-like objects (not necessarily null strings) 
which move with the speed
of light. The trajectories are specified by a set of spin coefficients introduced with respect to a tetrade $l,n,p,q$, where
$n,l$ are null, $l$ and $p$ are tangent to the trajectory, $l$ is the velocity of the string.  Rotations of the tetrade and reparametrizations
make a 2-parameter group of $l$-preserving null rotations of $n,p,q$, accompanied with
rescalings of $l$ and $n$. In general, the spin coefficients are not invariant with respect to this group. For null strings, however,
two spin-coefficients along the trajectories, $\theta_s=(p\cdot \nabla_p l)$ and $\kappa_2=(q\cdot \nabla_p l)$, are invariant up to boost rescalings.  In Sec. \ref{opt} we demonstrate that the complex quantity $Z=\theta_s+i\kappa_2$ satisfies an equation analogous
to Sachs' optical equations for null geodesic congruences (NGC). We call equation for $Z$ the string optical equation. In fact, Sachs' equations taken at the trajectory follow from string equations and vice versa. $Z$ is a linear combination
of complex divergence $\rho$ and complex shear $\sigma$ of NGC where string trajectory belongs to.  Scalars $\theta_s$ and $\kappa_2$ determine expansion and rotation 
of the string congruence.  Trajectory of the string in a space of parameters $\theta_s,\kappa_2$, the $Z$-plane, looks as a sequence 
of diagrams, see Sec. \ref{Pres}.   Examples of null strings in flat and  cosmological backgrounds and their diagrams 
are presented 
in Sec. \ref{sol}.  
In Sec.  \ref{BSF} we discuss null strings in asymptotically flat space-times. 
We consider strings on outgoing null hypersurfaces,  and, in particular, strings in the the Bondi-Sachs formalism. 
We show that the leading
asymptotics of $Z$ capture the amplitude of the outgoing gravitational radiation.  The gravitational memory effect in $Z$ caused by an ingoing flux of energy
is calculated for string trajectories in the weak field approximation.  Short discussion of our results can be found in Sec.
\ref{sum}.

\section{String trajectories}\label{def}
\setcounter{equation}0
\subsection{Spin coefficients and symmetries}

The `trajectory' of a string in a space-time $\cal M$ with coordinates $x^\mu$ is defined as $x^\mu=x^\mu(\lambda, \tau)$. Two real parameters, 
$\lambda$ and $\tau$ numerate points on the string. It is implied that the string trajectory has no caustics, or, in the case of caustics,
the definitions below are not applied at their locations.  One can introduce the tangent vectors 
$l^\mu\equiv x^\mu_{~,\lambda}$ and  $\eta^\mu \equiv x^\mu_{~,\tau}$.  For a string which moves with the speed of light we require 
that
\begin{equation}\label{1.1}
(l \cdot l )=0~~, 
\end{equation}  
\begin{equation}\label{1.2}
(l \cdot \eta)=0~~, 
\end{equation} 
where notation $(u \cdot v)$ stands for the scalar product in the tangent space of $\cal M$. We also assume that $l$ is future-directed, $\eta$ is space-like, 
$(\eta \cdot \eta )>0$. The velocity of the string is directed along $l$.  

Parameters $(\tau,\lambda)$ can be denoted as $\chi^a$, $a=1,2$.
The matrix $h_{ab}=x^\mu_{~,a}g_{\mu\nu}x^\nu_{~,b}$ is degenerate, 
$\det{h}=0$.  One can demand that $\det{h}=0$ at first. Then  (\ref{1.2}) follows from (\ref{1.1}), or  (\ref{1.1}) follows from (\ref{1.2}) .

The strings which are considered in this work obey (\ref{1.1}),  (\ref{1.2}) and the condition that each point 
of the string at fixed $\tau$ moves along a null geodesic, 
\begin{equation}\label{1.2a}
\nabla_l l\sim l~~. 
\end{equation} 
To put it another way, the string is a 1-parameter family of rays. Such a definition was first given by Schild \cite{Schild:1976vq}, and 
the corresponding strings are called `null strings'.

Trajectories of null strings can be considered as a certain type of null geodesic congruences (NGC).  
NGC play an important role in general relativity. They have been extensively studied since the mid of fifties of the last century.
The material of this Section follows in part the classical monograph  \cite{Penrose:1986ca}, where  $\eta$ is called the connecting vector between two rays.  Given (\ref{1.2})  a pair of of neighbouring rays are called abreast.  Eq. (\ref{1.2}) ensures that 
two points of the string with neighbouring world-lines always lie in 2-plane orthogonal to the worldlines.  

To study congruences which obey (\ref{1.1}),  (\ref{1.2}) we introduce a tetrade $l,n,p,q$ at the each point of the string trajectory. 
Here $p$ is a unit vestor directed along $\eta$, $p=\eta/N$, $N^2=(\eta \cdot \eta)$. Vector $n$ is null, orthogonal to $p$ and normalized as
$(n\cdot l)=-2$. Vector $q$ is spacelike, unit, and orthogonal to $l,n,p$. Note that the tetrade
cannot be introduced at fixed point sets of $\eta$, where $\eta=0$, and $N=0$. These points are caustics, where the interval
between two neighbouring rays vanishes.  We discuss these cases later, see
Sec. \ref{flat}. From now on we assume that $N\neq 0$.

The tetrade is defined up to null rotations (Lorentz
transformations of the parabolic type)
of $n$, $q$: 
\begin{equation}\label{1.3}
n=n'+2\omega q'+\omega^2 l'~~, ~~q=q'+\omega l'~~,
\end{equation} 
where $\omega$ may vary along the trajectory. The transformations leave invariant the tangent vectors $l=l'$ and $p=p'$. 

There is an arbitrariness in the choice of $l$ and $p$. Conditions (\ref{1.1}), (\ref{1.2}) allow 
reparametrizations
\begin{equation}\label{1.4}
\lambda'=g(\lambda,\tau)~~,~~\tau'=\phi(\tau)~~ 
\end{equation} 
which change tangent vectors to
$l'$, $\eta'$,
\begin{equation}\label{1.5}
l=g_{,\lambda} l'~~,~~\eta=\phi_{,\tau}\eta'+g_{,\tau} l'~~.
\end{equation} 
Then (\ref{1.5}) require change of $p$ and $n$ of the tetrade
\begin{equation}\label{1.3b}
n={1 \over g_{,\lambda}}(n'+2\bar{\omega} p'+\bar{\omega}^2 l')~~,~~p=p'+\bar{\omega} l'~~,~~\bar{\omega}={g_{,\tau} \over  N}~~,
\end{equation} 
similar to (\ref{1.3}). Reparametrizations (\ref{1.4}) generate null rotations of $p$ and $n$ and rescalings of $l$ and $n$.

Therefore, the tetrade  $l,n,p,q$  is defined up to a 2-parameter family of $l$-preserving null rotations of $n,p,q$ and 
rescalings of $l$ and $n$. These are class I and class III 
transformations, respectively, according to \cite{Chandrasekhar:1985kt}. We call (\ref{1.3})-(\ref{1.3b}) `$S$-transformations' , for  brevity.
Our aim is to construct from  $l,n,p,q$ quantities which are $S$-transform invariant.

We define $X_\lambda \equiv l$, $X_\tau \equiv \eta$, and denote reparametrizations (\ref{1.4}) as $(\chi')^a=(\chi')^a(\chi)$.
One can introduce a set of vectors $\nabla_a X_b$ ($\nabla_a=X_a^\mu \nabla_\mu$)  
tangent to the string world-sheet.   Since $\nabla_l \eta=\nabla_\eta l$, 
\begin{equation}\label{1.6}
\nabla_a X_b=\nabla_b X_a~~.
\end{equation}
$\nabla_a X_b$ yield 3 independent vectors which can be decomposed  as
\begin{equation}\label{1.7}
\nabla_a X_b=f^c_{ab}X_c+\kappa^n_{ab} n+\kappa^q_{ab} q~~.
\end{equation}
Coefficients  $f^c_{ab}$, $\kappa^n_{ab}$, $\kappa^q_{ab}$ are symmetric with respect to permutation of $a$ and $b$, and 
are related to spin coefficients for the chosen tetrade.  

With the help of (\ref{1.1}) and (\ref{1.2}) one finds
\begin{equation}\label{1.8}
\kappa^n_{\tau\tau}=\frac 12 N_{,\sigma}N~~,~~\kappa^n_{\tau\lambda}=\kappa^n_{\lambda\lambda}=0~~,
\end{equation}
\begin{equation}\label{1.9}
f^\tau_{\tau\tau}=\partial_\tau(\ln N)~~,~~f^\tau_{\lambda\tau}=\partial_\lambda (\ln N)~~,~~f^\tau_{\lambda\lambda}=0~~.
\end{equation}
The rest six coefficients, $f^\lambda_{ab}$, $\kappa^q_{ab}$, are not all independent and get mixed under $S$-transforms.  Null rotations (\ref{1.3})
yield relations
\begin{equation}\label{1.11}
\kappa^q_{ab}= (q\cdot \nabla_a X_b)=(\kappa')^q_{ab}-2\omega (\kappa')^n_{ab}~~,
\end{equation}
\begin{equation}\label{1.10}
f^\lambda_{ab}=-\frac 12 (n\cdot \nabla_a X_b)=(f')^\lambda_{ab}-\omega (\kappa')^q_{ab}+\omega^2(\kappa')^n_{ab}~~.
\end{equation}
Reparametrizations (\ref{1.4}) imply that
\begin{equation}\label{1.13}
\kappa^q_{ab}=(\chi')^{a'}_{~,a}(\chi')^{b'}_{~,b} (\kappa')^q_{a'b'}~~,
\end{equation}
\begin{equation}\label{1.12}
f^\lambda_{ab}= {1 \over g_{,\lambda}}(\chi')^{a'}_{~,a}(\chi')^{b'}_{~,b}
\left((f')^\lambda_{a'b'}-{g_{,\tau} \over \phi _{,\tau}}(f')^\tau_{a'b'}+{g^2_{,\tau} \over N^2}(\kappa')^n_{a'b'}\right)+
{1 \over g_{,\lambda}}\left(\lambda'_{,ab}-{g_{,\tau} \over \phi _{,\tau}}\tau'_{,ab}\right)~~,
\end{equation}
where we used (\ref{1.3b}) and (\ref{1.7}).

\subsection{Null strings and boost-weighted scalars}\label{sim}

A boost-weighted scalar (a $b$-scalar) along the string 
trajectory is defined as a scalar $Q$ which changes under $S$-transformations (\ref{1.3})-(\ref{1.3b}) as $Q=(g_{,\lambda})^bQ'$. Parameter $b$  is called 
the boost-weight of $Q$. Boost-weighted and spin-weighted quantities  are discussed 
in \cite{Penrose:1986ca} in the context of spin-coefficient formalism. 

The $b$-scalars play a key role in our analysis since they allow one to construct physical quantities measured 
by specific observers. In a frame of reference related to observers with velocities $u_o$ ($u^2_o=-1$) one can introduce
$b$-scalar $(u_o\cdot l)$ which is non-vanishing, since $u_o$ is time-like, and has boost-weight $b=1$.
If $Q$ is the b-scalar then $Q_o=(u_o\cdot l)^{-b}Q$ is S-transform invariant scalar, $b=0$. In the given frame, $Q_o$ can be interpreted as a physical observable.

We construct $b$-scalars from  $\kappa^q_{ab}$.
Coefficients in (\ref{1.8}), (\ref{1.9}) depend only on $N$.   Parameter $N$ determines length
of a small segment of the string, $dL=N d\tau$.  $S$-transformations are $N=\phi_{,\tau}N'$, $dL=dL'$.  One can introduce 
the expansion parameter
\begin{equation}\label{1.14}
\theta_s=\partial_\lambda (\ln N)=(p\cdot \nabla_p l)~~,
\end{equation}
which is $b=1$ scalar. $\theta_s$ measures how fast $dL$ changes along the world-line of a point on the string, $\partial_\lambda (dL)=\theta_s (dL)$. 
The next set of scalars is related to $\kappa^q_{ab}$.  It follows from (\ref{1.11}), (\ref{1.13}) and (\ref{1.5}) that 
\begin{equation}\label{1.16}
\kappa_1\equiv \kappa^q_{\lambda\lambda}~~
\end{equation}
is $b=2$ scalar. According to (\ref{1.7}), (\ref{1.8}), (\ref{1.9}) 
\begin{equation}\label{1.17}
\nabla_l l= f^\lambda_{\lambda\lambda} l+\kappa_1 q~~.
\end{equation}
Hence, condition $\kappa_1=0$ implies that the string is null. 

Coefficient $\kappa^q_{\lambda\tau}$ does not change under null rotations (\ref{1.11}) but transforms
as   
\begin{equation}\label{1.18}
\kappa^q_{\lambda\tau}=g_{,\lambda} \phi_{,\tau}(\kappa')^q_{\lambda'\tau'}+g_{,\lambda}g_{,\tau}(\kappa')^q_{\lambda'\lambda'}~~
\end{equation}
under reparametrizations  (\ref{1.5}). If the string is null the last term in the r.h.s. of  (\ref{1.5}) is zero and 
\begin{equation}\label{1.19}
\kappa_2\equiv (q \cdot \nabla_p l )=N^{-1}\kappa^q_{\lambda\tau}~~,
\end{equation}
is weight $b=1$ scalar. Scalars $\kappa_2$ and $\theta_s$ play an important role in the subsequent analysis.  We show that $\kappa_2$ measures 
rotation of $\eta$ in a plane orthogonal to the velocity of the string under a parallel transport of $\eta$ along world-line, see Sec. \ref{modes}.

One can continue in this way to come to the following $b$-scalars  under certain restrictions.
\begin{equation}\label{1.20}
\kappa_3 \equiv (q \cdot \nabla_p p ) =N^{-2}\kappa^q_{\tau\tau}~~
\end{equation}
is $b=0$ scalar, if $\theta_s=\kappa_1=\kappa_2=0$.
When the string is not null ($\kappa_1\neq 0$), 
\begin{equation}\label{1.24}
\kappa_4 \equiv N^{-2} \det \kappa^q_{ab}~~
\end{equation}
is $b=2$ scalar, if $\theta_s=0$.  

Coefficients $f^\lambda_{ab}$ yield no scalars due to the last term in the r.h.s. of (\ref{1.12}).

Note that $\theta_s$ and $\kappa_1$ are the $b$-scalars which do not require any conditions. Other $\kappa_i$  require vanishing of spin coefficients.  Condition 
$\kappa_1=0$ can be imposed in any space-time: null strings, 
like rays,  can always be constructed. Hence, $\kappa_2$ can be introduced for null strings.

In  certain space-times, null strings include
a subclass of strings with $\theta_s=\kappa_2=0$. Then $\kappa_3$ can be considered as a physical parameter. This subclass includes
subclass with $\theta_s=\kappa_2=\kappa_3$.

 Strings in space-times with global parabolic isometries \cite{Fursaev:2017aap}, \cite{Fursaev:2018spa}, 
\cite{Fursaev:2020oom} are of special interest since they allow explicit description of backreaction effects.
They were called massless strings.  The world-sheets of such strings are null 2-surfaces which are
fixed points sets of null rotations. The metric of a space-time which allows a global parabolic isometry is \cite{Fursaev:2020oom} 
\begin{equation}\label{1.30}
ds^2=-2e^\lambda (dudv-dy^2)+h(dz+s du)^2~~, 
\end{equation} 
where $\lambda$, $h$ and $s$ are functions of $u$, $z$ and $\theta=uv-y^2$.
The isometries of (\ref{1.30}) are null rotations of coordinates
\begin{equation}\label{1.32}
u'=u~~,~~
v'=v+2\omega y+\omega^2u~~,~~
y'=y+\omega u~~,~~z'=z~~,
\end{equation} 
which leave $\theta$ invariant. 
The trajectory of a massless string is given by simple equations:  $u=y=0$. One can check that it fullfills (\ref{1.1}),  (\ref{1.2}), (\ref{1.2a}).
That is the string is null.

Now let $\zeta$ be the Killing field which generates (\ref{1.32}) and the corresponding null rotation
of the tetrade (\ref{1.3}). Since $\zeta=0$ on the world-sheet one gets the set of conditions:
\begin{equation}\label{1.26}
0=\delta_\zeta (\nabla_a X_b\cdot n)=(\nabla_a X_b\cdot \delta_\zeta n)=2\omega (\nabla_a X_b\cdot q)=2\omega \kappa^q_{ab}~~,
\end{equation}
\begin{equation}\label{1.27}
0=\delta_\zeta (\nabla_a X_b\cdot q)=(\nabla_a X_b\cdot \delta_\zeta q)=\omega (\nabla_a X_b\cdot l)=-2\omega \kappa^n_{ab}~~.
\end{equation}
By taking into account (\ref{1.8}) one concludes that strings studied in \cite{Fursaev:2017aap}, \cite{Fursaev:2018spa}, 
\cite{Fursaev:2020oom} are null strings for which $\theta_s=\kappa_1=\kappa_2=\kappa_3=0$.

\section{Optical equations}\label{opt}
\setcounter{equation}0
\subsection{String scalar and its equation}

Consider now relation between spin coefficients at a string trajectory and the curvature of the background space-time $\cal M$.
It follows from (\ref{1.6}) that
\begin{equation}\label{2.1}
[\nabla_a ,\nabla_b] X^\mu_c=-R^\mu_{~\alpha \rho\nu}X^\alpha_c X^\rho_a X^\nu_b,
\end{equation}
where $R^\mu_{~\alpha \rho\nu}$ is the Riemann tensor of $\cal M$ at the string trajectory (we use 
definition $R^\mu_{~\alpha \rho\nu}V^\alpha \equiv [\nabla_\nu ,\nabla_\rho] V^\mu$).  The left hand side of  (\ref{2.1}) can 
be calculated with the help of (\ref{1.7}). After some algebra, replacing $\eta$ with $p$ one gets a set of relations
for components of the Riemann tensor at the string trajectory
\begin{equation}\label{2.2}
R_{p lp l}=(\partial_l -\beta_1)\theta_s+\theta_s^2+\kappa_1\kappa_3-\kappa_2^2~~,
\end{equation}
\begin{equation}\label{2.3}
R_{q l p l}=(\partial_l-\beta_1) \kappa_2+2\theta_s \kappa_2-(\partial_p-\beta_2) \kappa_1~~,
\end{equation}
\begin{equation}\label{2.4}
R_{q p lp}=(\partial_p+\beta_2) \kappa_2-(\partial_l+\theta_s) \kappa_3-\beta_3 \kappa_1+\frac 12\theta_s \alpha_1~~,
\end{equation}
\begin{equation}\label{2.5}
R_{n l l p }=(\partial_l+2\theta_s) \beta_2-\kappa_2\alpha_1+\kappa_1\alpha_2~~,
\end{equation}
\begin{equation}\label{2.6}
R_{n p l p }=2\left((\partial_l+\theta_s) \beta_3-\partial_p \beta_2+\beta_3\beta_1-\beta_2^2\right)+\kappa_2\alpha_2-\kappa_3\alpha_1~~.
\end{equation}
We put
$\beta_1= - \frac 12 (n \cdot \nabla_l l ) =f^\lambda_{\lambda\lambda}$, $\beta_2= - \frac 12 (n \cdot \nabla_p l )$, $\beta_3 = - \frac 12 (n \cdot \nabla_p p )$, 
$\alpha_1=(\nabla_l q\cdot n)$, $\alpha_2=(\nabla_p q\cdot n)$, and use notation 
$R_{abcd}=R_{\mu\alpha \rho\nu}e^\mu_a e^\alpha_b e^\rho_c e^\nu_d$ for $e_a=l,p,n,q$.

Eqs. (\ref{2.2}), (\ref{2.3}) are of the most interest since $R_{p lp l}$, $R_{q l p l}$ are $b=2$ scalars.  One can use these
equations to  get an analog of Raychaudhuri type equations and draw physical information.
Other curvatures  in (\ref{2.4})-(\ref{2.3})
are not boost weighted scalars, in general.
Under $S$-transformations $R_{q p lp}$, $R_{n l l p }$ get mixed with $R_{p lp l}$, $R_{q l p l}$. Thus,
$R_{q p lp}$, $R_{n l l p }$ are $b=2$ and $b=1$ scalars, respectively, only if  $R_{p lp l}=R_{q l p l}=0$. 
Analogously $R_{n p l p }$ is $b=1$ scalar,
if $R_{p lp l}=R_{q l p l}=R_{n l l p }=0$.  

From now on we assume that the string is null, $\kappa_1=0$. 

It is convenient to introduce a pair of null complex vectors $\hat{m}$, 
$\hat{\bar{m}}$, $(\hat{m}\cdot \hat{\bar{m}})=1$,
\begin{equation}\label{2.7}
\hat{m}={1 \over \sqrt{2}}(p+iq)~~,~~\hat{\bar{m}}={1 \over \sqrt{2}}(p-iq)~~.
\end{equation}
By following the standard procedure \cite{Penrose:1986ca}, \cite{Chandrasekhar:1985kt} one defines invariants
$C_{abcd}$ and $R_{ab}$ constructed from the Weyl tensor and Ricci tensor, respectively, 
$$
\hat{\Psi}_0=-C_{\hat{m}l\hat{m}l}~~,~~\Phi_{00}=-\frac 12 R_{ll}~~.
$$
Since $C_{\hat{m}l\hat{\bar{m}}l}=0$, see \cite{Chandrasekhar:1985kt}, it follows that
\begin{equation}\label{2.8}
R_{p lp l}=C_{p lp l}+\frac 12 R_{ll}=-\mbox{Re}~\hat{\Psi}_0-\hat{\Phi}_{00}~~,
\end{equation}
\begin{equation}\label{2.9}
R_{q lp l}=C_{q lp l}=-\mbox{Im}~\hat{\Psi}_0~~.
\end{equation}
We also introduce complex $b=1$ scalar
\begin{equation}\label{2.10}
Z=\theta_s+i\kappa_2= ((p+iq)\cdot \nabla_p  l)~~.
\end{equation} 
With these definitions Eqs. (\ref{2.2}), (\ref{2.3}) in the case of null strings take the simple form:
\begin{equation}\label{2.11}
D_lZ+Z^2=-\hat{\Psi}_0-\Phi_{00}~~,
\end{equation}
where $D_l\equiv \partial_l-\beta_1$ is a covariant derivative with respect to boost transformations. 

We call $Z$ the string scalar. This parameter plays an important 
role for the rest of the article.   By taking square of the left and and right hand sides of (\ref{1.7}) for $a=\tau, b=\lambda$ 
one gets a useful identity
\begin{equation}\label{2.11a}
(\nabla_\eta l \cdot \nabla_\eta l )=N^2(\theta_s^2+\kappa_2^2)~~,
\end{equation}
where we took into account (\ref{1.9}). Then it follows from (\ref{2.10}) that
\begin{equation}\label{2.11b}
|Z|=|\nabla_p l |~~.
\end{equation}
We call (\ref{2.11}) optical equation for strings, by analogy with optical equations of Sachs for NGC. 
The importance of (\ref{2.11}) is that it can be used to draw some universal features of string trajectories, for example, in 
asymptotically flat or asymptotically de Sitter space-times, where curvatures  $\hat{\Psi}_0$, $\Phi_{00}$ decay fast enough at null infinities.

\subsection{Relation to Sachs' equations}

To establish relation between (\ref{2.11}) and Sachs equations we introduce another complex dyad at the string trajectory, $m$ and $\bar{m}$,
\begin{equation}\label{2.12}
(m\cdot \bar{m})=1~~,~~(m\cdot l)=(m\cdot n)=0~~.
\end{equation}
The set $(n,l,m,\bar{m})$ is the doubly null tetrade.
Note that $m$ and $\bar{m}$ are arbitrary vectors while dyad $\hat{m}$, 
$\hat{\bar{m}}$ is connected with $\eta$ and $q$, and implies the condition $\nabla_l\eta=\nabla_\eta l$.
The fact that $(n,l,m,\bar{m})$ are not restricted by any conditions allows one to use
the Newman-Penrose formalism  \cite{Penrose:1986ca}, \cite{Chandrasekhar:1985kt} and require that $m$ is parallel transported along the
world-line, $\nabla_l m=0$.
We put $\hat{m}=N\zeta^{-1}m$, where $\zeta$ is a complex parameter, $|\zeta|=N$.
The connecting vector in the  new basis is
\begin{equation}\label{2.13}
\eta={1 \over \sqrt{2}}\left(\zeta \bar{m}+\bar{\zeta} m\right)~~.
\end{equation}
When a point moves along the world-line the phase of $\zeta$ determines orientation of
$\eta$ in the plane $(m,\bar{m})$. The condition $\nabla_l\eta=\nabla_\eta l$ requires 
that 
\begin{equation}\label{2.14}
\partial_l \zeta=-\rho \zeta -\sigma \bar{\zeta}~~,
\end{equation}
\begin{equation}\label{2.15}
\rho \equiv-(m\cdot \nabla_{\bar{m}} l)~~~,~~\sigma \equiv-(m\cdot \nabla_m l)~~.
\end{equation}
Here we took into account that  $\nabla_l m=0$. Spin coefficients  $\rho$, $\sigma$ are known
as optical scalars which are defined for NGC with velocity vector $l$.  One can check that $\rho$, $\sigma$  are $b=1$ boost-weighted scalars on trajectories of null strings. It follows from definition (\ref{2.10}) that
\begin{equation}\label{2.16}
\zeta Z=-\zeta\rho -\bar{\zeta}\sigma~~.
\end{equation}
So (\ref{2.14}) implies 
\begin{equation}\label{2.17}
\partial_l \zeta= Z \zeta~~.
\end{equation}
If string optical equation (\ref{2.11}) is satisfied one can take the derivative $D_l$  of the left and right parts 
of (\ref{2.16}), use (\ref{2.14}), (\ref{2.17}) to get
\begin{equation}\label{2.18}
-\zeta \Phi_{00}-\bar{\zeta}\Psi_0=\zeta(\rho^2+|\sigma|^2-D_l\rho)+\bar{\zeta}(\sigma(\rho+\bar{\rho})-D_l\sigma)~~,
\end{equation}
\begin{equation}\label{2.19}
\Psi_0=-C_{mlml}=\bar{\zeta}^{-1}\zeta \hat{\Psi}_0~~.
\end{equation}
Eq. (\ref{2.18}) requires the following relations known as optical equations of Sachs:
\begin{equation}\label{2.20}
D_l\rho=\rho^2+|\sigma|^2+\Phi_{00}~~,
\end{equation}
\begin{equation}\label{2.21}
D_l\sigma=\sigma(\rho+\bar{\rho})+\Psi_0~~.
\end{equation}
Sachs equations (\ref{2.20}), (\ref{2.21}) are derived for a general NGC and are not related to string trajectories.
Given a string trajectory, Sachs equations at the trajectory may follow from (\ref{2.11}). Opposite is true as well: if a string trajectory is a
one-parameter family of rays in a NGC, then string equation (\ref{2.11})  follows from
Sachs equations for the given NGC.

\section{Representation of string trajectories}\label{Pres}
\setcounter{equation}0

\subsection{Diagram description of string trajectories}\label{fix}

The physical meaning of $\theta_s$, the real part of the string scalar $Z$, is related to local expansion (contraction) of string segment 
moving along the world-line, see Sec. \ref{sim}. To find interpretation of $\kappa_2$, the imaginary part of $Z$, we
use (\ref{2.17}). When a point of the string moves along the world-line the components $(\zeta,\bar{\zeta})$ of the connecting vector 
with respect to dyad $(m, \bar{m})$ change,  
\begin{equation}\label{3.1}
\zeta(\lambda+\delta\lambda)\simeq (1+Z\delta \lambda)\zeta(\lambda)~~.
\end{equation}
If $\zeta=|\zeta|e^{i\alpha}$, Eq. (\ref{3.1}) implies change of the phase 
\begin{equation}\label{3.2}
\alpha(\lambda+\delta\lambda)=\alpha(\lambda)+\mbox{Im}~Z ~\delta \lambda~~.
\end{equation}
Therefore, $\mbox{Im}~Z$ determines rotation of $\eta$ in the plane $(m, \bar{m})$. The rotation
angle of $\eta$ under the shift $\delta\lambda$ is $\kappa_2\delta\lambda$.  The sign of the rotation is connected with 
the sign of $\kappa_2$. It changes if $q$ is replaced to $-q$. This reflection arbitrariness can be eliminated by additional 
arguments. For example, one can introduce a null vector  $\tilde{l}_\mu=\epsilon_{\mu\nu\lambda\rho}p^\nu q^\lambda l^\rho$.
Since $\tilde{l}=a l$ one can fix the sign of $q$ by requiring, for example, that $a>0$.

In next sections we study constant $\lambda$ slices of the string world-sheet. These slices are curves which
determine shape of the string. 
If $\lambda$ is fixed, $Z=Z(\lambda,\tau)$ is a curve in a space of parameters $\theta_s,\kappa_2$, or in a 
complex $Z$-plane.  String scalar 
$Z$ can be used to describe evolution of the shape of the string in a given slicing. 

We call such curves
diagrams of the string trajectory. The diagrams can be written  locally
as $\theta_s=\theta_s(\kappa_2)$ or $\kappa_2=\kappa_2(\theta_s)$, so they do not depend on spatial parametrization
of the trajectory by $\tau$.  String diagrams yield ``portraits'' of the string trajectories. 

\begin{figure}[h]
\begin{center}
\includegraphics[height=7cm,width=14.0cm]{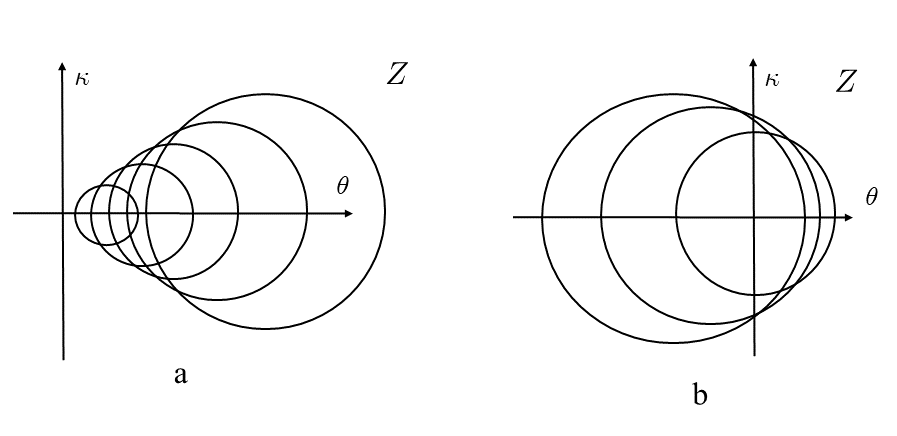}
\caption{\small{string diagrams for string scalar $Z$ given by (\ref{2.22}) with $z=ce^{i\tau}$. 
For $\lambda>r$, $(a)$, circles shrink as $\lambda$ grows.  For $0<\lambda<r$, $(b)$, circles
expand as $\lambda$ grows.}}
\label{f3} 
\end{center}
\end{figure}

To construct the string diagrams one needs to remember
that $\lambda$ is not uniquely defined and allows transformations (\ref{1.5}). 
One can use physical arguments and 
relate $\lambda$ to a frame of reference where the string trajectory is considered.
If $u_o$ are 4-velocities of observers which make a certain frame of reference, then it is natural 
to require that
\begin{equation}\label{3.3}
(u_o\cdot \eta)=0~~.
\end{equation}
Condition (\ref{3.3}) fixes $\lambda$ up to rescalings $\lambda'=g(\lambda)$ which leave constant $\lambda$ slices invariant
and, hence, do not change string diagrams.
The rescalings can be restricted  by further arguments. For instance, in some cases one can require that $(u_o\cdot l)=-1$.
Examples of string trajectories with such conditions are presented in Secs. \ref{flat1}, \ref{cosm}. 

\subsection{Exact solutions for $Z$}\label{flat1}

To give an idea of the string diagrams  consider strings in space-times where $\hat{\Psi}_0=\Phi_{00}=0$. One can choose 
affine paramerization, $D_l=\partial_l$, to get a general solution to (\ref{2.11})
\begin{equation}\label{2.22}
Z(\lambda,\tau)={1 \over \lambda+z(\tau)}~~,
\end{equation}
where $z$ is a complex function.  Scalars $\theta_s$ and $\kappa_2$ are real and imaginary parts of the r.h.s. of (\ref{2.22}), respectively.
Solutions like (\ref{2.22}) hold for strings in conformally-flat space-times with $\Phi_{00}=0$. Eq. (\ref{2.22}) holds for strings in de Sitter and anti-de Sitter geometries.  Singularities of $Z$ at $\lambda=-z(\tau)$ may appear in different models. They correspond to caustics
at fixed points of the vector field $\eta$, see below.

As an example, consider the case when $z=ce^{i\tau}$, where $c$ is a positive constant. One finds
$|\lambda|\neq c$:
\begin{equation}\label{2.22b}
\theta_s(\lambda,\tau)={\lambda +c\cos\tau \over (\lambda + c\cos\tau)^2+c^2\sin^2\tau}~~,~~
\kappa_2(\lambda,\tau)=-{c\sin\tau \over (\lambda +c \cos\tau)^2+c^2\sin^2\tau}~~,
\end{equation}
which is equivalent to
\begin{equation}\label{2.22c}
(\theta_s-d)^2+\kappa_2^2=R^2~~,
\end{equation}
\begin{equation}\label{2.22d}
d=d(\lambda)={\lambda \over \lambda^2-c^2}~~,~~R=R(\lambda)={c \over |\lambda^2-c^2|}~~.
\end{equation}

\begin{figure}[h]
\begin{center}
\includegraphics[height=4.3cm,width=16.0cm]{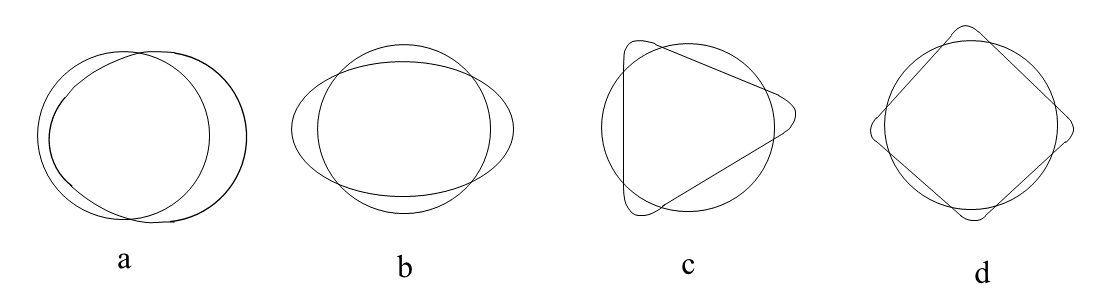}
\caption{\small{shows schematically first 4 deformation modes of a circular string in the $(p,q)$ 
plane when $\lambda$ is slightly increased. Deformations $a$,$b$,$c$,$d$ correspond to
$n=1,2,3,4$. }}
\label{f1} 
\end{center}
\end{figure}

One can represent $Z$ as
\begin{equation}\label{2.22e}
Z(\lambda,\tau)=d(\lambda)+R(\lambda)e^{i\xi} ~~,
\end{equation}
\begin{equation}\label{2.22f}
e^{i\xi}=\pm {c+ \lambda e^{i\tau} \over \lambda + c e^{i\tau}}~~,
\end{equation}
where signs $+$ and $-$ corresponds to the cases $\lambda<c$ and $\lambda>c$, respectively. Eq. (\ref{2.22f}) is a linear fractional
transformation from $z=e^{i\tau}$ to $w=e^{i\xi}$. Therefore, the diagram of the  string at fixed $\lambda$ is a circle.
The diagrams for $|\lambda|\neq c$ are shrinking or expanding circles shown on Fig. \ref{f3}.

The case $\lambda=c$ is special, 
$$
Z(c,c)={e^{-i\tau/2} \over 2c\cos{\tau \over 2}}~~.
$$
Let us emphasize that the diagrams depend on the choice of $\lambda$. 
Examples given above are for $\lambda$ being an affine parameter.
The residual freedom of changing $\lambda$ to $\lambda'=\lambda+f(\tau)$ can be eliminated by additional conditions like
(\ref{3.3}). Examples are given in Sec. \ref{flat}.

\subsection{Closed strings}\label{modes}

To see how parameters $\theta_s$ and $\kappa_2$ determine transformation
of the shape of the string consider, as an illustration, a closed string,  $Z(\lambda,\tau+2\pi)=Z(\lambda,\tau)$.
Suppose that $\lambda$-slices are fixed by (\ref{3.3}).

One can use (\ref{1.3}) to choose vector $q$ orthogonal to $u_o$. Then the pair $(p,q)$ yield  
a 2-plane in the frame of reference of the observers. If one considers the Fourier transform
\begin{equation}\label{3.4}
Z(\lambda,\tau)=\sum_n e^{in\tau}c_n(\lambda)~~,
\end{equation}
coefficients $c_n$ determine different  transformations of the shape of the string, the string modes. Assume that only the single
mode $n\neq 0$ is present in (\ref{3.4}) and $c_n>0$ .  
The corresponding expansion and rotation scalars are
\begin{equation}\label{3.5}
\theta_s(\lambda,\tau)=c_n(\lambda)\cos n\tau~~,~~\kappa_2(\lambda,\tau)=c_n(\lambda)\sin n\tau~~.
\end{equation}
At fixed $\lambda$,   parameters $\tau= 2 k \pi/n$ ,  $k=0,1,2...$, are the points of maximal expansion,
$\theta_s>0$,  while maximal contraction occurs at $\tau= (2 k+1)\pi/n$, $\theta_s<0$. There is no rotation of $\eta$ at these points, $\kappa_2$=0.  
Between $\tau= 2 k \pi/n$ and $\tau= (2 k+1)\pi/n$ rotation of $\eta$ is, say, counter-clockwise, $\kappa_2>0$,  from  $\tau= 2 k \pi/n$ to $\tau= (2 k+1)\pi/n$.
Between $\tau= (2 k-1) \pi/n$ and $\tau= 2k \pi/n$ rotation is clockwise, $\kappa_2<0$, from  $\tau= 2 k \pi/n$ to $\tau= (2 k-1)\pi/n$.
That is, points of the string rotate toward a nearby point of maximal contraction.

String modes have simple form if the string is a circle. String modes $n=1,2,3,4$ are shown on Fig. \ref{f1}.

\section{Examples of string trajectories}\label{sol}
\setcounter{equation}0

\subsection{Strings in Minkowsky space-time}\label{flat}

A general solution to Eqs.   (\ref{1.1}) - (\ref{1.2a}) for a null string in a flat space-time, for
the choice of $\lambda$ as an affine parameter, 
is
\begin{equation}\label{4.1}
X^\mu(\lambda,\tau)=\lambda b^\mu(\tau)+a^\mu(\tau)~~,
\end{equation}
where $b^\mu$ is an arbitrary null vector, $b^2=0$. Restrictions on $a^\mu$ are: $(b\cdot \dot{a})=0$,
 see (\ref{1.2}), and $\dot{a}^2>0$, $\dot{a}\equiv a_{,\tau}$.  One finds:
\begin{equation}\label{4.2}
N^2=\lambda^2 \dot{b}^2+2\lambda (\dot{b} \cdot \dot{a})+\dot{a}^2~~,
\end{equation}
\begin{equation}\label{4.3}
\theta_s=N^{-2}(\lambda \dot{b}^2+ (\dot{b} \cdot \dot{a}))~~.
\end{equation}
To calculate the rotation scalar we use (\ref{2.11a}) 
\begin{equation}\label{4.4}
\kappa_2^2=N^{-2}\dot{b}^2-\theta_s^2=N^{-4}\left(\dot{a}^2\dot{b}^2-(\dot{b} \cdot \dot{a})^2\right)~~.
\end{equation}
In the flat space-time we can fix parametrization of the string-world sheet for an inertial frame of reference by conditions
$(l\cdot u_o)=-1$, $(\eta \cdot u_o)=0$. In Minkowsky coordinates, where velocity of observers is $u_o^\mu=\delta^\mu_0$, these
conditions are ensured if $t(\lambda,\tau)=\lambda$, that is $\dot{a}$ and $\dot{b}$ have only spatial components.
Let $|\dot{b}|\neq 0$, and
\begin{equation}\label{4.5}
\cos\varphi\equiv {(\dot{b} \cdot \dot{a}) \over |\dot{a}||\dot{b}|}~~,~~r\equiv { |\dot{a}| \over |\dot{b}|}
\end{equation}
Then under appropriate choice of the sign of $\kappa_2$ the string scalar is
\begin{equation}\label{4.6}
Z(\lambda,\tau)={1 \over \lambda+r(\tau) e^{i\varphi (\tau)}}~~,
\end{equation}
in accord with (\ref{2.22}). The rotation scalar vanishes when $\dot{a}=0$, or $\varphi=0$.

Let us consider several examples.

\bigskip

{\bf 1}.  Take Eqs. (\ref{4.1}) in Minkowsky coordinates as $t=x=\lambda, y=y(\tau), z=z(\tau)$. The corresponding string 
moves along the $x$-axis, the string lies in a 2-plane orthogonal  to the direction of motion.  
One can check that $Z\equiv 0$. The string diagrams are a single dot, since the strings do not change its shape. 

\bigskip

\begin{figure}[h]
\begin{center}
\includegraphics[height=8cm,width=8.0cm]{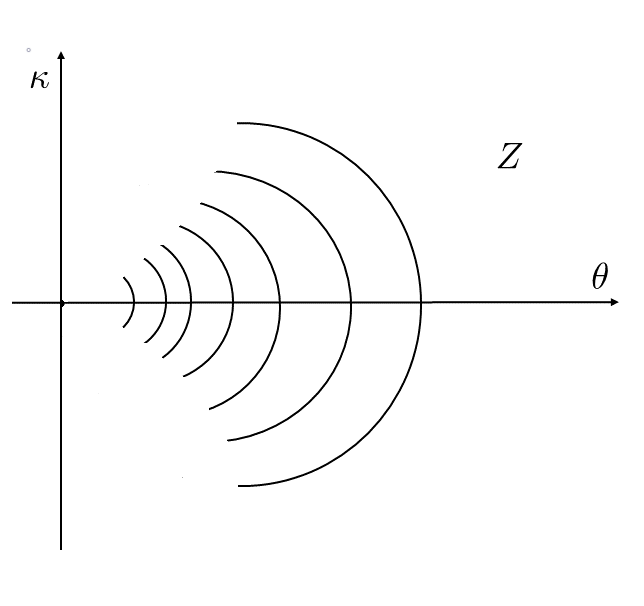}
\caption{\small{diagrams of a closed string in Minkowsky space-time. The diagrams are arcs of
circles that shrink as $\lambda$ grows.  }}
\label{f2} 
\end{center}
\end{figure}

{\bf 2}.  An interesting string trajectory is:
\begin{equation}\label{4.6ab}
t=\lambda~~,~~x=\lambda\cos\tau+\frac c4\cos 2\tau~~, ~~y=\lambda \sin\tau+\frac c4(\sin 2\tau+2\tau)~~, ~~z=c\cos\tau~~,
\end{equation} 
where $c>0$ is a constant. At fixed $\lambda$ the string is twisted along $y$ axis.  It is instructive
to calculate the corresponding connecting vector field $\eta^\mu=\dot{x}^\mu$:
\begin{equation}\label{4.6ac}
\eta^t=0~~,~~\eta^x=-\sin\tau(\lambda+ c\cos\tau)~~, ~~\eta^y=\cos\tau(\lambda+ c\cos\tau)~~, ~~\eta^z=-c\sin\tau~~.
\end{equation} 
This vector field vanishes, $\eta=0$, at two fixed points of the congruence: $\lambda=r,\tau=\pi$ and  $\lambda=-r,\tau=0$. 
One can check that 
\begin{equation}\label{4.6ad}
b^2=(b\cdot \dot{a})=0~~,~~\dot{b}^2=1~~,~~r=|\dot{a}|=c~~,~~(\dot{b} \cdot \dot{a})=c\cos\tau=c\cos\varphi~~.
\end{equation} 
Therefore, $\theta_s$ and $\kappa_2$ in the $Z$-plane satisfy (\ref{2.22c}), (\ref{2.22d}), for $\lambda\neq r$.
$Z$ has simple poles at fixed points of $\eta$. 

The diagrams of the string are shrinking or expanding circles shown on Fig. \ref{f3}.

\bigskip

{\bf 3}.  Consider a closed string described by
equations
\begin{equation}\label{4.6b}
t=\lambda~~, ~~x=\lambda \cos\tau~~, ~~y=\lambda \sin\tau~~, ~~z=c\sin\tau~~,
\end{equation}
where $c>0$ is a constant. For such string 
\begin{equation}\label{4.6e}
b^2=(b\cdot \dot{a})=0~~,~~\dot{b}^2=1~~,~~r=|\dot{a}|=c|\cos\tau|~~~,~~(\dot{b} \cdot \dot{a})=0~~.
\end{equation} 
\begin{equation}\label{4.6c}
Z(\lambda,\tau)={1 \over \lambda+ic\cos\tau}~~.
\end{equation}

It follows from (\ref{4.6c})  that $\theta_s$ and $\kappa_2$ are related as
\begin{equation}\label{4.6d}
(\theta_s-R)^2+\kappa_2^2=R^2~~~~R={1 \over 2\lambda}~~.
\end{equation}
\begin{equation}\label{4.6f}
Z(\lambda,\tau)=R(\lambda)+R(\lambda)e^{i\xi} ~~,
\end{equation}
\begin{equation}\label{4.6g}
e^{i\xi}= {\lambda - ic \cos\tau \over \lambda + ic \cos\tau}~~.
\end{equation}
Map  (\ref{4.6g}) from $z=e^{i\tau}$ to $w=e^{i\xi}$ is not a linear fractional transformation.  At fixed $\lambda$ string portrait is only an arc of the circle (\ref{4.6d}). 
Ends of the arc are at points $\xi=\pm \xi_0$, where $\partial_\tau \xi=0$. This corresponds to $\tau=0,\pi$, and 
$$
\sin\xi_0={2\lambda c \over \lambda^2+c^2}~~.
$$
Each point on the arc corresponds to two values of $\tau$. The diagrams of the string are shown on Fig. \ref{f2} for positive $\lambda$.  

In the above examples string diagrams have a common feature: the diagrams shrink to point $Z=0$ as $\lambda$ grows. At large $\lambda$, or as 
null future infinity is approached, all deformations of string's shape gradually decay. One can say that the null strings are ``freezing out''.

\subsection{Strings in cosmological models}\label{cosm}

Studying null cosmic strings in exрanding universe is of a particular interest, since such objects may result
in observable physical effects \cite{Fursaev:2018spa}.  We consider conformally flat cosmologies with
\begin{equation}\label{4.15}
ds^2=-dt^2+a^2(t)(dx^i)^2~~,
\end{equation}
where scale factor $a(t)$ is determined by a concrete model.  Coordinate $t$ is the cosmological time. 
The string equations are defined as
\begin{equation}\label{4.17}
t=\lambda~~,~~x^i(\lambda,\tau)=f(\lambda) b^i(\tau)+a^i(\tau)~~,
\end{equation}
\begin{equation}\label{4.18}
\partial_\lambda f = a^{-1}(\lambda)~~,
\end{equation}
where $(b^i)^2=1$, $b^i\dot{a}^i=0$. The tangent vectors are $l=\partial_t+f'~b^i\partial_i$, $\eta=(f~ \dot{b}^i+\dot{a}^i)\partial_i$.
With the help of (\ref{4.18}) one checks that 
\begin{equation}\label{4.15a}
l^2=0~~,~~\nabla_l l=\chi l~~,~~(\eta \cdot l)=0~~,
\end{equation}
where $\chi=\partial_\lambda a/a$. Thus, (\ref{4.17}) describe a null string.
In the frame of freely
moving observers with 4-velocities $u_o^\mu=\delta^\mu_t$  the following conditions hold:
\begin{equation}\label{4.16}
(u_o\cdot l)=-1~~,~~(u_o\cdot \eta)=0~~.
\end{equation}
Eq. (\ref{4.16}) can be used to fix the parametrization.  Relation between $\lambda$ and the affine parameter $\bar{\lambda}$ is 
$\partial_\lambda\bar{\lambda}=C(\tau)a(\lambda)$, where $C(\tau)$ is an arbitrary real function.

A straightforward calculation yields
\begin{equation}\label{4.19}
Z(\lambda,\tau)={1 \over a(\lambda)}{1 \over f(\lambda)+z(\tau)}+\partial_\lambda \ln a~~,
\end{equation}
where $z(\tau)=r(\tau) e^{i\varphi (\tau)}$ is defined by (\ref{4.5}).  For a universe filled with a matter
with the equation of state $p=w\rho$, $a\sim \lambda^p$, $f\sim \lambda^{1-p}$, where
$p=2/(3(1+w))$.  One concludes that for dust or radiation dominated universe  $Z(\lambda,\tau)=O(\lambda^{-1})$ at large $\lambda$. 
This means that strings do not change their shape at large cosmological time $t$.

For a flat de Sitter universe, $a(\lambda)=e^{H\lambda}$,
\begin{equation}\label{4.23}
Z(\lambda,\tau)={H^2z(\tau)  \over H z(\tau)-e^{-H\lambda}}~~.
\end{equation}
At large $\lambda$, $Z(\lambda,\tau)\simeq H$.  Cosmic strings at late times do not rotate
but expand exponentially, similarly to other scales.

Scalar $Z$  vanishes, if  $a^i$ is a constant, $z=0$ in (\ref{4.23}). Trajectories 
of these strings lie on the cosmological 
horizon $|x^i(t)-a^i|=H^{-1}e^{-Ht}$. A particular type of such strings, massless strings on the equator of the horizon sphere, admit 
exact analysis of backreaction effects, and has been studied in detail in \cite{Fursaev:2018spa}.

String diagrams for (\ref{4.17}) can be constructed analogously to the case of string trajectories in Minkowsky space-time.

\section{Strings in asymptotically flat space-times}\label{BSF}
\setcounter{equation}0

\subsection{Strings on null hypersurfaces}\label{null}

String trajectories may lie on null hypersurfaces. Then $l$ is directed along a normal vector to the surface. This case 
can be studied by using the Bondi-Sachs formalism. 
One can always choose a tetrade basis on the entire space-time
so that it coincides with $n,l,\hat{m},\hat{\bar{m}}$ defined at the string trajectory, see Sec. \ref{opt}, and introduce corresponding spin-coefficients. 
As we saw, equation (\ref{2.16}) relates the string scalar $Z$ to the spin coefficients $\rho$ and $\sigma$ which are divergence and 
shear of a null geodesic congruence. In the chosen basis, $\zeta=1$ in (\ref{2.16}).

This fact is important since  asymptotic form of the shear at future null infinity $\cal{I}^+$ together with
asymptotics of the complex tetrade components of the Weyl  tensor $\Psi_k$ are used to extract interior physical properties of the space-time,
see  \cite{Adamo:2009vu}, for a review. We return to this issue in Sec. \ref{gw}.

In asymptotically flat space-times curvature scalars behave as $\Psi_0=O(\lambda^{-5})$, 
$\Phi_{00}=O(\lambda^{-6})$ at $\cal{I}^+$. Therefore the r.h.s. of (\ref{2.11}) can be
ignored and (\ref{2.22}) can be used to get for expansion and rotation scalars the following asymptotics
at large $|\lambda|$:
\begin{equation}\label{4.7}
Z\simeq{1 \over \lambda}-{z(\tau)\over \lambda^2}+...~~,~~\theta_s\simeq{1 \over \lambda}-{z_1(\tau)\over \lambda^2}+..~~,~~\kappa_2\simeq -{z_2(\tau) \over \lambda^2}+...~~,
\end{equation} 
where $z_1=\mbox{Re}~ z$, $z_2=\mbox{Im}~ z$.  We see from (\ref{4.7}) the already familiar phenomenon: shapes of strings in asymptotically flat space-times 
cease to vary at large $\lambda$, as $\cal{I}^+$ is approached. Possible ripples on the strings are encoded in the subleading term in $Z$, and are
determined in (\ref{4.7}) by a complex parameter $z(\tau)$.

Asymptotic behavior of a NGC on an outgoing
null  hypersurface is \cite{Adamo:2009vu}
\begin{equation}\label{7.1}
\rho=\bar{\rho}\simeq-{1 \over \lambda}-{\sigma^0\bar{\sigma}^0 \over \lambda^3}+...~~,
~~
\sigma\simeq {\sigma^0 \over \lambda^2}+..~~,
\end{equation}
where $\sigma^0$ is called the asymptotic complex shear of the NGC. If the string belongs to the given NGC and the tetrades are chosen
so that they coincide with $n,l,\hat{m},\hat{\bar{m}}$ at the string trajectory, one can use (\ref{2.16}), to find the simple
relation 
\begin{equation}\label{7.2}
z=\sigma^0~~.
\end{equation}
That is, the subleading term in the string scalar $Z$ is determined by the asymptotic shear $\sigma^0$ calculated in the special basis.

Note that the string trajectory itself may be forming a null 2 surface. This surface is defined by conditions $f=0$, $t=0$, and $l_\mu=a f_{,\mu}$,
$q_\mu=b t_{,\mu}$, where $a$ and $b$ are set on the surface. Then $(q\cdot \nabla_q l)=-\partial_l\ln b=0$, and one gets an additional
relation
\begin{equation}\label{1.25}
\theta \equiv \nabla_ll=(p\cdot \nabla_p l)+(q\cdot \nabla_q l)-\frac 12 (n\cdot \nabla_l l)=\theta_s~~.
\end{equation}
For a general NGC with velocity $l$ parameter $\theta $ measures expansion or contraction of the area of 
2d space-like sections of the congruence. Thus, if the NGC is null surface forming, the string trajectory belongs to NGC and it is
2-surface forming, then the string expansion $\theta_s$ coincides with the expansion of NGC.

\subsection{Strings in the Bondi-Sachs formalism and gravitational waves}\label{gw}

The asymptotic shear $\sigma^0$ is known to measure the amplitude of gravity waves far from the source. We now
find out an explicit relation between asymptotic form of $Z$, see (\ref{4.7}), and outgoing gravitational radiation. 
We consider strings in asymptotically flat space-times, far from a source of gravity waves.  It is 
convenient to use the Bondi-Sachs coordinates $x^\mu=(u,r,x^A)$, $A=1,2$, based on a family of outgoing null hypersurafces, 
see \cite{Madler:2016xju} for a review. 
The corresponding metric is
\begin{equation}\label{5.1}
ds^2=-Udu^2-2e^{2\beta}dudr+g_{AB}(dx^A- V^Adu)(dx^B-V^Bdu)~~.
\end{equation}
The null hypersurfaces in question are  $u=c$, where $c$ is a constant. Coordinate $r$ which varies along null rays is chosen to be areal coordinate.  
The future null  infinity is at $r\to+\infty$.

Metric (\ref{5.1})
is flat when $U=1$, $\beta=V^A=0$, $g_{AB}=r^2\gamma_{AB}$, with $\gamma_{AB}$ being a metric on a unit 2-sphere.
In an asymptotically flat space-time at large $r$
\begin{equation}\label{5.2}
U=1-{2M \over r}+O(r^{-2})~~,
\end{equation}
\begin{equation}\label{5.2b}
g_{AB}=r^2\left(\gamma_{AB}+{C_{AB} \over r}+O(r^{-2})\right)~~,
\end{equation}
Other parameters, $\beta$ and $V^A$, are $O(r^{-2})$, see \cite{Madler:2016xju}.  The parameter $M=M(u,x^A)$ is the mass aspect. 
$C_{AB}=C_{AB}(u,x^A)$ is a traceless tensor in a tangent space to $S^2$. The term ${C_{AB} / r}$ in (\ref{5.2b}) 
is a perturbation of the metric caused by the outgoing gravitational radiation.

Vector $l=\partial_r=-e^{-2\beta}\nabla u$ is a tangent vector to null geodesics on the null hypersurfaces $u=c$. We choose the ``gauge''
$\beta=0$. Then $\nabla_l l=0$, and $r$ can be identified with an affine parameter.  The string equations in coordinates
$u,r,x^A$  are
\begin{equation}\label{5.3}
u=c~~,~~r=\lambda~~,~~x^A=x^A(\tau)~~.
\end{equation}
The connecting vector is $\eta=\eta^A\partial_A$, where $\eta^A=\dot{x}^A$. One can check with the help of (\ref{5.1}) that (\ref{1.1}),  (\ref{1.2}) are satisfied and 
\begin{equation}\label{5.4}
\theta_s={1 \over 2}p^A p^B\partial_r g_{AB}~~,~~p^A={\eta^A \over (\eta^A\eta^B g_{AB})^{1/2}}~~.
\end{equation}
The simplest choice for the vector $q$ in the tetrade at the string trajectory is  $q=q^A\partial_A$,
\begin{equation}\label{5.5}
q^A=g^{AC}\epsilon_{CB}p^B~~,~~\epsilon_{AB}=-\epsilon_{BA}~~,~~\epsilon_{12}=(\det g_{AB})^{1/2}~~.
\end{equation}
$A,B$ are risen and lowered with the help of $g_{AB}$ and its inverse matrix. (\ref{5.5}) is not a unique choice but it is 
enough to find $\kappa_2$. A straightforward calculation with the help of (\ref{1.19}), (\ref{5.1}) yields
\begin{equation}\label{5.6}
\kappa_2={1 \over 2}q^A p^B\partial_r g_{AB}~~,
\end{equation}
\begin{equation}\label{5.7}
Z(\lambda,\tau)={1 \over 2}(p^A+iq^A) p^B\partial_r g_{AB}~~.
\end{equation}
We are interested in asymptotic properties of $Z$ at large $r=\lambda$, which easily follow from (\ref{5.2}),  
$$
\partial_r g_{AB}={2g_{AB} \over r} -C_{AB}+O(r^{-1})~~,
$$
\begin{equation}\label{5.8}
Z(\lambda,\tau)={1 \over \lambda}-{z \over \lambda^2} +O(\lambda^{-3})~~,
\end{equation}
\begin{equation}\label{5.9}
z={1 \over 2}(\bar{p}^A+i\bar{q}^A) \bar{p}^B C_{AB}~~.
\end{equation}
Here we took into account that $p^A=\bar{p}^A/r+O(r^{-2})$, $q^A=\bar{q}^A/r+O(r^{-2})$. In fact,
$\bar{p}^A$, $\bar{q}^A$ are unit mutually orthogonal vectors tangent to $S^2$
\begin{equation}\label{5.10}
\bar{p}^A={\eta^A \over (\eta^A\eta^B \gamma_{AB})^{1/2}}~~,~~\bar{q}_A=\bar{\epsilon}_{AB}\bar{p}^B~~,~~\bar{\epsilon}_{12}=(\det \gamma_{AB})^{1/2}~~.
\end{equation}
Since the metric can be written as $\gamma_{AB}=\bar{p}_A\bar{p}_B+\bar{q}_A\bar{q}_B$,  and $C_{AB}$ is traceless, $C_{AB}\gamma^{AB}=0$, 
one can decompose
\begin{equation}\label{5.11}
C_{AB}=(\bar{p}_A\bar{p}_B+\bar{q}_A\bar{q}_B)C_++(\bar{p}_A\bar{q}_B+\bar{q}_A\bar{p}_B)C_{\mbox{x}}~~,
\end{equation}
and rewrite (\ref{5.9}) in a simple form
\begin{equation}\label{5.12}
z=-{1 \over 2}(C_++iC_{\mbox{x}})~~.
\end{equation}
Coefficients $C_+$ and $C_{\mbox{x}}$ correspond to ``$+$'' and ``$\mbox{x}$'' polarizations of a gravity wave in the given basis. 
Expression (\ref{5.12}) is in accord with the fact that ``$+$'' and ``$\mbox{x}$'' polarizations are related to real and imaginary parts of the shear.

\subsection{Interaction of the string with ingoing flux}\label{aflat}

Consider an outgoing null string which moves toward $\cal{I}^+$ and an ingoing flux of matter which crosses the string trajectory.
Our aim is to find out how the flux affects parameter $z(\tau)$ in asymptotic (\ref{4.7}). 
We assume that the flux is weak and  its interaction with all points of the string happens in a short interval 
$(\lambda_\star, \lambda_\star+\Delta \lambda)$.
 
Substitution $Z=Y^{-1}$ brings (\ref{2.11}) to the form:
\begin{equation}\label{4.8}
\partial_\lambda Y=1+{\cal R}Y^2~~,~~{\cal R}=\hat{\Psi}_0+\Phi_{00}~~,
\end{equation}
in the affine parametrization.  We assume that ${\cal R}$ is entirely due to the flux.  Solution to (\ref{4.8}) can be found perturbatively,
\begin{equation}\label{4.9}
Y=Y_0+Y_1+Y_2+...~~,
\end{equation}
\begin{equation}\label{4.10}
\partial_\lambda Y_0=1~~,~~\partial_\lambda Y_1={\cal R}Y_0^2~~,
\end{equation}
where $Y_k=O({\cal R}^k)$. Solutions to (\ref{4.9}), (\ref{4.10}) are
\begin{equation}\label{4.11}
Y_0(\lambda,\tau)=\lambda+z(\tau)~~,~~ Y_1(\lambda,\tau)=\int^\lambda_{\lambda_\star}{\cal R}(\tau,\lambda')(\lambda'+z(\tau))^2d\lambda'~~.
\end{equation}
Before interaction with the flux, $\lambda < \lambda_\star$ 
\begin{equation}\label{4.11a}
Z(\lambda,\tau)\simeq {1 \over \lambda}-{z(\tau)\over \lambda^2}~~.
\end{equation}
After the interaction, at $\lambda> \lambda_\star+\Delta \lambda$, (\ref{4.9}), (\ref{4.11}) imply that
\begin{equation}\label{4.12}
Z(\lambda,\tau)\simeq {1 \over Y_0} -{Y_1 \over Y^2_0}\simeq {1 \over \lambda}-{z(\tau)+\Delta z\over \lambda^2}~~,
\end{equation}
\begin{equation}\label{4.13}
\Delta z(\tau)\simeq  {\cal R}(\tau,\lambda_\star)\lambda_\star^2 \Delta \lambda~~.
\end{equation}
The shift $\Delta z$ is the gravitational memory left after the interaction with the flux. 

In the Einstein gravity, if $\Psi_0=0$, the memory effect is due to ${\cal R}=-4\pi G  T_{ll}$, where $T_{ll}$ is the null component  
of the stress-energy tensor of ingoing matter.  
In this case  $\Delta z$ is real, and the flux does not cause additional rotation of the string.

\section{Summary}\label{sum}

The aim of this paper was to develop a coordinate, parametrization and basis independent description of null strings.
We identified a complex spin coefficient $Z$, which is a boost-weighted  scalar. $Z$
obeys a string optical equation and determines transformation of the string along its trajectory. 

The string diagrams introduced in the $Z$-plane may be a useful tool to study the string trajectories in simple terms.
The diagrams are portraits of the strings. Some examples of the diagrams 
have been presented for strings in flat space-times and in conformally flat cosmological models.  Constructing diagrams 
for other physically interesting
situations, such as null strings interacting with black holes or null strings in realistic
cosmological models, is left for further research.

Null cosmic strings, as tensile strings, may result in a number of observable effects in their environment. The 
strings create overdensities of matter, additional anisotropy of CMB, and etc. All these effects
are related to a backreaction of the geometry caused by strings.
 
In the present paper we were interested in the opposite effects: how geometry affects null strings, and, in particular,
asymptotic properties of their trajectories.
It follows from the string optical equation 
that all transformations of the shape of strings in asymptotically flat space-times gradually decay. Strings are ``freezing out'' as the future null infinity 
$\cal{I}^+$ is approached. The same property holds for null strings in  cosmological models of flat dust or matter dominated universes.
In asymptotically de Sitter space-times strings stop to rotate but stretch accordingly with the cosmological expansion.

Interactions of null strings with the background curvature and flows of matter 
cause ripples on the strings near $\cal{I}^+$. These gravitational memory effects are encoded in the subleading terms of $Z$.
If null cosmic strings are fundamental tensionless strings produced in the early universe and stretched to cosmological scales the 
ripples may carry an important information about the Planckian physics.  This would be an intriguing feature of null
cosmic strings, analogous to features of relic gravitational waves, and another research topic.

\end{document}